\documentclass{article}
\usepackage{graphicx}
\usepackage{amssymb}
\usepackage{amsmath}
\usepackage{enumitem}
\usepackage[utf8]{inputenc}
\usepackage{esvect}
\usepackage[left=3cm,right=3cm,top=3cm,bottom=3cm]{geometry}
\usepackage{authblk}
\usepackage{etoolbox}
\usepackage{color}

\makeatletter
\patchcmd{\@maketitle}{\raggedright}{\centering}{}{}
\patchcmd{\@maketitle}{\raggedright}{\centering}{}{}
\makeatother

\begin{document}

\title{Nonlocal Quantum Field Theory and Quantum Entanglement}
\author{Robin Landry$^{\dag,\ddag}$ \& John W. Moffat$^{\dag,*}$}
\affil{$^{\dag}$Perimeter Institute for Theoretical Physics, Waterloo, Ontario N2L 2Y5, Canada}
\affil{$^{\ddag}$Département de Physique, Université Paris-Saclay, Orsay, F91405, France}
\affil{and}
\affil{$^{*}$Department of Physics and Astronomy, University of Waterloo, Ontario N2L 3G1, Canada}
\date{July 2023}
\maketitle

\begin{abstract}
    We discuss the nonlocal nature of quantum mechanics and the link with relativistic quantum mechanics such as formulated by quantum field theory. We use here a nonlocal quantum field theory (NLQFT) which is finite, satisfies Poincaré invariance, unitarity and microscopic causality. This nonlocal quantum field theory associates infinite derivative entire functions with propagators and vertices. We focus on proving causality and discussing its importance when constructing a relativistic field theory.  We formulate scalar field theory using the functional integral in order to characterize quantum entanglement and the entanglement entropy of the theory. Using the replica trick, we compute the entanglement entropy for the theory in $3+1$ dimensions on a cone with deficit angle. The result is free of UV divergences.

\end{abstract} \textit{robin.landry@universite-paris-saclay.fr\\ jmoffat@perimeterinstitute.ca}

\tableofcontents

\newpage
\section{Introduction}
\label{sec:introduction}

Quantum entanglement is one of the most bizarre features of quantum mechanics. When two quantum systems interact and become entangled, their quantum states are correlated in a non-classical way. Measuring a property of one system seems to instantaneously influence the other, even if they are separated by a large distance. Bell’s theorem states that the predictions made by quantum mechanics, concerning correlations between different measurements performed on physically separate systems cannot be reproduced by any local hidden variable theory, because these predictions are in fact incompatible with Bell’s inequality~\cite{bell,einstein}. The violation of Bell's inequality has been experimentally verified, separately by Aspect, Clauser and Zeilinger~\cite{A1, A2, C1, C2, Z}.

Quantum field theory is a more fundamental theory to explain the microscopic world than quantum mechanics. One of the most fundamental aspects of the microscopic world and which is highlighted by quantum mechanics is nonlocality, that is to say the existence of quantum correlations between entangled systems.

Quantum mechanics is an inherently nonlocal theory that violates causality. Nonlocal quantum field theories appear in various form~\cite{pais,efimov,alebastrov,krasnikov,rev,M2,M3,M1,M5,M6,M4,M7,M8,nortier1,Modesto2,Modesto3,Koshelev,Chin,Tomboulis}. We will explore a nonlocal quantum field theory (QFT), formulated as a scalar field theory. In this formulation, nonlocality is an inherent property of the field operators. For a field taken at a point x, the field depends on the entire spacetime. This is achieved by regularizing the fields by an entire function distribution operator. The nonlocal quantum field theory satisfies microscopic causality~\cite{M5,M6}.

The primary motivation of this formulation did not originally reside in the nonlocality of its operators, but in the fact that this nonlocal regularization makes the theory finite to all orders of perturbation theory~\cite{rev,M2}. There is therefore no need to involve infinite renormalization, since the ultraviolet divergences do not intervene. The nonlocal regularization of quantum field theory preserves Poincaré invariance, unitarity and microcausality. The problem of infinite renormalizability is resolved by regulated propagators and vertices for the Feynman loop diagrams of QED, QCD and weak interactions~\cite{M5}. The propagators are regulated by an infinite derivative entire function $\mathcal{E}(p^2)$, which is analytic and holomorphic in the complex $p^2$ plane with a pole and/or an essential singularity at infinity. This finite Poincaré invariant and unitary QFT allows us to formulate the standard model of particle physics~\cite{M6} and can be extended to quantum gravity~\cite{M7,M8}. This choice of a finite theory is therefore to be considered as a potential candidate for a more complete formulation of QFT~\cite{M5,M6}.

In order to study quantum entanglement, one of the most meaningful ways is to study entanglement entropy. This entropy quantifies the quantum entanglement between entangled systems. By calculating it for our quantum field theory, we can therefore observe the direct consequences of taking nonlocality as a founding hypothesis for quantum field theory. To access the entanglement entropy of the scalar field theory, we have to determine the partition function of the theory. The easiest way to do this is to use the path integral formalism.

Currently, quantum entanglement is explained in the framework of algebraic quantum field theory (AQFT), a local theory in the sense of operator algebras, and quantum entanglement is explained via the Reeh-Schlieder theorem~\cite{AQFT}. It implies that by acting on the vacuum with any local operator supported in a small region, we can create whatever state we wish in a spacelike separated region. Whereas the entanglement entropy is always UV divergent in standard local scalar field theory, we show that it is finite using a nonlocal scalar field theory. And because it is the nonlocal assumption that brings finiteness in this theory, we interpret quantum correlations between entangled systems, i.e quantum entanglement, as a direct consequence of nonlocality.

In the first part of this paper, we discuss the nonlocal interpretation of quantum mechanics. We define quantum entanglement and review Bell's inequality. In the second part of the paper, we explore the consequences of taking a nonlocal quantum field theory to describe quantum entanglement. We propose a nonlocal scalar QFT as an example of a causal nonlocal theory. We then show that it satisfies microscopic causality. Finally, we implement the theory using the path integral formalism, which allows us to calculate the entanglement entropy in the theory in $3+1$ dimensions on the cone with deficit angle.

\newpage

\section{Quantum Mechanics, Quantum Entanglement and Nonlocality}
\label{sec:QM}

\subsection{Quantum entanglement}

In quantum mechanics, in order to define quantum entanglement, we have to consider the pure state of a composite system A and B :

\begin{equation}
    |\psi \rangle \in \mathcal{H}_A \otimes \mathcal{H}_B.
\end{equation} If this pure state can be written in the form of a tensor product between a state A and a state B,

\begin{equation}
    |\psi \rangle = |\phi \rangle_A \otimes |\phi \rangle_B,
\end{equation} then $|\psi \rangle$ is a separable state and is therefore not entangled.

Whereas quantum entanglement is the mathematical formulation of the inseparability of certain quantum states, nonlocality describes the physical correlations between two separable states. Nonlocality implies quantum entanglement and nonlocality is an interesting path to study quantum entanglement in quantum field theory.

One way to learn whether two systems are entangled is through entanglement entropy. Entanglement entropy makes it possible to quantify the entanglement of a composite system AB. It is a quantity that vanishes, if and only if the composite system AB is not entangled.

The entanglement entropy is the von Neumann entropy of the reduced state of each subsystem A and B. One interesting and useful property of entanglement entropy is that the von Neumann entropy of each subsystem of a composite system is the same. For a composite system AB :

\begin{equation}
    S = -Tr(\rho_A\ln(\rho_A)) =  -Tr(\rho_B\ln(\rho_B)),
\end{equation} where $\rho_A$ and $\rho_B$ are the reduced density matrices of the subsystems A and B.

The calculation of this quantity is a non-trivial problem in quantum field theory. We will see that it is more convenient to opt for a functional integral approach by reformulating the entropy of entanglement in terms of the partition function of the theory.

\subsection{Bell's theorem}

Bell's theorem states that the predictions made by quantum mechanics, concerning correlations between different measurements performed on physically separate systems cannot be reproduced by any local hidden variable theory, because these predictions are in fact incompatible with Bell's inequalities~\cite{bell,einstein}.

Bell's inequalities show that the principle of locality dictates that correlations between different measurements performed on physically separate systems must satisfy certain conditions. Bell's research highlighted the conditions imposed on quantum mechanics by local causality. These Bell inequalities group together a number of inequalities reproduced by local hidden variable theories.

Experiments ~\cite{A1, A2, C1, C2, Z} have verified the violation of Bell's inequalities, excluding local theories of hidden variables and focus on the nonlocal nature of quantum mechanics.
\newpage
\section{Nonlocal Quantum Field Theory}
\label{sec:nonlocal}

We adopt here the Minkowski metric convention $\eta_{\mu\nu}=$ diag($+1$,$-1$,$-1$,$-1$), and we set $\hbar = c = k_B = 1$.

\subsection{Nonlocal Scalar Field Theory}

Let us consider a nonlocal interacting scalar field theory, which describes a spin$-0$ particle with a mass $m$. It is the scalar version of the more general particle physics Nonlocal Quantum Field Theory (NLQFT) developed in~\cite{M1,M5,M6,M4}. The lagrangian density of our theory is given by

\begin{equation}
    \mathcal{L} =  \frac{1}{2}\partial_\mu\tilde{\phi}\partial^\mu\tilde{\phi} - \frac{1}{2}m^2\tilde{\phi}^2 - \frac{\lambda}{4!}\tilde{\phi}^4,
\end{equation} where we define the nonlocal field operator

\begin{equation}
    \tilde{\phi}(x) = \int d^4x' \mathcal{F}(x - x')\phi(x') = \mathcal{F}(\Box_x)\phi(x).
\end{equation}
In this expression, $\phi(x)$ denotes the local field operator and $\Box_x=\partial_\mu\partial^\mu$.

The regularized position propagator $\tilde{\Delta}(x - x')$ in Minkowski spacetime is the Green's function $G(x, x')$ for the Klein-Gordon equation~\cite{M5}:

\begin{equation}
    (\square_x + m^2)\tilde{\Delta}(x - x') \equiv \mathcal{E}(x - x') = -\frac{1}{4\pi \Lambda_x^4}\exp\left(-(x - x')^2/2\Lambda_x^2\right).
\end{equation} We note that taking the limit $\Lambda_x \to 0$, we find the local Klein-Gordon equation :

\begin{equation}
    (\square_x + m^2)\Delta(x - x') = -\delta^4(x - x'),
\end{equation} where $\Delta(x - x')$ is the local position propagator.

We have replaced the local dynamics of fields by a nonlocal one. The field at one point of spacetime depends on all other positions in spacetime. By using entire function distribution operators our theory is made to be finite, which means there is no need for infinite renormalization, because the theory is free of UV-divergences.

\subsection{Causality}

Let us consider the transition amplitude from $x$ to $x'$ for two spacelike separated points in spacetime. For a free particle $\langle x'|{\rm e}^{-iHt}|x \rangle$, one can show that the amplitude is non-zero in the case of non-relativistic quantum mechanics as well as in relativistic quantum mechanics. In nonrelativistic quantum mechanics we have $E = \frac{\vv{p}^2}{2m}$ and obtain~\cite{peskin}:

\begin{equation}
    \langle \vv{x}'|{\rm e}^{-i\frac{p^2}{2m}t}|\vv{x} \rangle = \left(\frac{m}{2\pi it}\right)^{\frac{3}{2}}{\rm e}^{im\frac{(\vv{x} - \vv{x}')^2}{2t}}.
\end{equation} The transition amplitude is nonzero, which means that a particle can propagate between two points in an arbitrarily short time. For two points outside the light cone, this leads to a violation of causality.

Now let us consider the case of relativistic quantum mechanics where $E = \sqrt{\vv{p}^2 + m^2}$. Similar calculations as previously lead to a result expressed in term of Bessel functions. By looking at the asymptotic behavior outside the light cone : $x^2 \gg t^2$, we obtain~\cite{peskin}:

\begin{equation}
    \langle \vv{x}'|{\rm e}^{-i\sqrt{\vv{p}^2 + m^2}t}|\vv{x} \rangle \sim {\rm e}^{-m\sqrt{\vv{x}^2 - t^2}}.
\end{equation} The transition amplitude is nonzero and therefore causality is violated. The introduction of creation and annihilation operators in manifestly Lorentz invariant QFT solves the causality problem and allows us to describe multiparticles states.

Let us review the proofs that that two nonlocal field operators $\tilde{\phi}(x)$ and $\tilde{\phi}(x')$ commute at spacelike separated points $x$ and $x'$~\cite{M5,M6}. We have
\begin{align}
[\tilde{\phi}(x),\tilde{\phi}(x')]&= \langle 0|[\tilde{\phi}(x),\tilde{\phi}(x')]|0\rangle\\
&= \tilde{\Delta}(x - x') - \tilde{\Delta}(x' - x).
\end{align}
where $\tilde{\Delta}(x - x')$ is the nonlocal propagator. Note that the two right-hand side terms are manifestly Lorentz invariant. Hence, by taking a spacelike spacetime interval, $(x - x')^2 < 0$ and by performing a Lorentz transformation on the second term on the right-hand side: $(x - x') \to  -(x - x')$, the two propagators become equal and the commutator vanishes. Micro-causality and therefore causality are preserved~\cite{peskin}. It means that two measurements cannot affect one another outside the light cone.

Let us consider an alternative proof of nonlocal microcausality~\cite{M6}. The local field operator satisfies the commutation relation
\begin{equation}
[\phi(x), \phi(x')]  = i\bar{\Delta}(x - x').
\end{equation} Here, $\bar{\Delta}(x - x')$
is the Pauli-Jordan propagator defined by
\begin{equation}
\bar{\Delta}(x - x') = \int \frac{d^4p}{(2\pi)^4}{\rm e}^{-ip\cdot(x - x')}\epsilon(p^0)2\pi\delta(p^2 - m^2),
\end{equation}
where
\begin{equation}
\epsilon(p^0) = \theta(p^0) - \theta(-p^0) = \begin{cases}
        +1 &\text{ if }p^0 > 0 \\
        -1 &\text{ if }p^0 < 0
    \end{cases},
\end{equation}
and

\begin{equation}
\theta(p^0) = \begin{cases}
        1 &\text{ if }p^0 > 0 \\
        0 &\text{ if }p^0 < 0
    \end{cases}.
\end{equation}

The Pauli-Jordan propagator vanishes outside the light cone:
\begin{equation}
[\phi(x), \phi(x')] = 0, \,\, (x - x')^2 < 0.
\end{equation}

We define~\cite{M6}:
\begin{equation}
{\cal F}(x-x')={\cal F}(\Box_x)\delta^4(x-x'),
\end{equation}
where
\begin{equation}
\tilde\phi(x)=\int d^4x'{\cal F}(x-x')\phi(k)\exp(-ik\cdot x)=\int\frac{d^4k}{(2\pi)^4}\phi(k){\cal F}(-k^2)\exp(-ik\cdot x).
\end{equation}
The nonlocal field operator commutator is given by
\begin{equation}
[\tilde\phi(x),\tilde\phi(x')]=[{\cal F}(\Box_x)\phi(x),{\cal
F}(\Box_{x'})\phi(x')]={\cal F}(\Box_x){\cal F}(\Box_{x'})[\phi(x),\phi(x')].
\end{equation}
We now obtain
\begin{align}
[\tilde\phi(x),\tilde\phi(x')]&={\cal F}(\Box_x){\cal F}(\Box_{x'})\int\frac{d^4k}{(2\pi)^4}\exp(-ik\cdot
(x-x'))\epsilon(k^0)2\pi\delta(k^2-m^2)\nonumber\\
&=\int\frac{d^4k}{(2\pi^4)}{\cal F}^2(-m^2)\exp(-ik\cdot
(x-x'))\epsilon(k^0)2\pi\delta(k^2-m^2)\\ &={\cal F}^2(-m^2)(-i)\bar\Delta(x-x').
\end{align}
For the nonlocal field operator $\tilde\phi(x)$, it follows that
\begin{equation}
    [\tilde\phi(x),\tilde\phi(x')]=0,\quad (x-x')^2 < 0.
\end{equation}
This proves that the nonlocal QFT satisfies microscopic causality.

We have demonstrated that the NLQFT is causal, no information is exchanged at a speed faster than light. The nonlocal nature of field operators due to entire function distribution operators at vertices and in propagators does not lead to a violation of causality.

\subsection{Path integral formulation of NLQFT}

In this section, we formulate our nonlocal scalar field theory in the path integral formalism. The path integral is based on the Lagrangian formalism and the relation between the Hamiltonian and the Lagrangian of our scalar field theory is the canonical one, allowing us to write the transition amplitude in the following way:

\begin{equation}
    \langle \tilde{\phi}(x_1), t_1|\tilde{\phi}(x_2), t_2 \rangle= \int \mathcal{D}\tilde{\phi} \; exp\left[i\int_{t_1}^{t_2} d^4x \left(
    \frac{1}{2}\partial_\mu\tilde{\phi}\partial^\mu\tilde{\phi} - \frac{1}{2}m^2\tilde{\phi}^2 -\frac{\lambda}{4!}\tilde{\phi}^4\right)\right].
    \label{path}
\end{equation}

We observe that taking the continuous limit in this finite, nonlocal theory the measure $\mathcal{D}\tilde{\phi}$ does not lead to UV divergences, because the theory is UV complete.

It is useful to consider the partition function $Z[\tilde{J}]$, which is the time ordered Green's function in the vacuum. Our goal by computing this partition function is to access the entanglement entropy of our nonlocal theory and $\tilde{J}$ is a nonlocal external source current. We use the generating functional method, which consists of adding a nonlocal source term $\tilde{J}(x)\tilde{\phi}(x)$ in the original path integral (\ref{path}) :

\begin{align}
    Z[\tilde{J}] &= \int \mathcal{D}\tilde{\phi} \; \exp\left[i\int_{t_1}^{t_2} d^4x \left(
    \frac{1}{2}\partial_\mu\tilde{\phi}\partial^\mu\tilde{\phi} - \frac{1}{2}m^2\tilde{\phi}^2 - \frac{\lambda}{4!}\tilde{\phi}^4 + \tilde{J}\tilde{\phi}\right)\right]\\
    &= Z_0[0]\exp\left[{\frac{-i\lambda}{4!}\int d^4x \tilde{\phi}^4}\right]\exp\left[{\frac{-i}{2}\int d^4x_1 d^4x_2 \tilde{J}(x_1)\tilde{\Delta}_F(x_1 - x_2)\tilde{J}(x_2)}\right].
\end{align} where $Z_0[0] = \langle 0|0 \rangle_{J = 0} = 1$ is the functional integral of the free field.

The nonlocal Feynman propagator in Euclidean position space is given by

\begin{equation}
    \label{prop}
    \tilde{\Delta}_F(x - x') = \int \frac{d^4p}{(2\pi)^4}\frac{{\rm e}^{-(p^2 + m^2)/2\Lambda_p^2}}{p^2 + m^2}{\rm e}^{-ip\cdot(x - x')},
\end{equation} where we have adopted the entire function distribution function ${\cal E}(p^2)=\exp(-(p^2+m^2)/2\Lambda_p^2)$. The UV completion of the nonlocal theory leads to a finite result for $\Delta_F(0)$, a notable difference with a local scalar field theory, where $\Delta_F(0)$ diverges quadratically in perturbation theory and requires an infinite renormalization to give a valid physical result.

Amplitudes in QFT can be defined in both Minkowski and Euclidean spaces, and they can be analytically continued from one signature to the other. In local QFT in which interaction vertices are local (defined in polynomials in momenta), the Feynman loop integrals converge to zero in the large loop momenta. The Cauchy theorem for infinite-radius semicircles allows the application of Wick rotation in the complex momentum space. However, in nonlocal QFT (nonpolynomial in momenta) the Minkowski and Euclidean momentum spaces are not equivalent. The transcendental entire functions that make loop integrals ultraviolet convergent can, at the same time, make integrands diverge along a complex direction when $|\exp(-f(p^2))|\rightarrow\infty$, in the limit $|p^0|\rightarrow\infty$ for an angle $\theta$ of the complex loop energy $p^0=|p^0|\exp^{i\theta}$. However, methods can be defined to deform integration contours consistent with the conditions of unitarity and analyticity~\cite{buoninfante}. We shall employ a functional path integral to calculate a partition function and the quantum entanglement entropy. A path integral is convergent only if it is defined initially in the Euclidean space signature. Moreover, in standard local QFT renormalization and power counting analysis is only valid in the Euclidean signature space. A successful prescription for nonlocal theories can only be formulated when the path integral and loop calculations are performed in Euclidean space. The calculated amplitudes and physically observable quantities depending on real momenta in nonlocal QFT can be obtained by an analytic continuation from Euclidean space to Minkowski space. The analytic continuation prescription requires that initial calculations of nonlocal QFT loops, amplitudes and partition functions are performed in Euclidean space and then analytically continued to Minkowski space according to a certain defined prescription~\cite{buoninfante}.

To obtain an analytical form for the entanglement entropy, we have to study further the partition function. As the scalar field theory does not have a closed form expression for $Z[\tilde{J}]$, we have to evaluate it perturbatively. We perform the following transformation~\cite{das}:

\begin{equation}
    \tilde{\phi}(x) \to \frac{\delta}{\delta \tilde{J}(x)}.
\end{equation} A first-order Taylor expansion in $\lambda$ gives

\begin{equation}
    Z[\tilde{J}] = Z_0[0]\left(1 - \frac{i\lambda}{4!}\int d^4x \frac{\delta^4}{\delta \tilde{J}^4(x)}\right)\exp{\left[\frac{-i}{2}\int d^4x_1 d^4x_2 \tilde{J}(x_1)\tilde{\Delta}_F(x_1 - x_2)\tilde{J}(x_2)\right]}.
\end{equation}
Let us consider the case with no external source current, $\tilde{J}(x) = 0$ :

\begin{equation}
    Z[0] = Z_0[0]\left(1 +\frac{i\lambda}{8}\int d^4x \tilde{\Delta}_F(0)\tilde{\Delta}_F(0)\right).
\end{equation} By choosing the spherical polar coordinates $\phi, \theta, \chi, \kappa$, we obtain from (\ref{prop}):

\begin{align}
    \tilde{\Delta}_F(0) &= \frac{1}{(2\pi)^4}\int_{0}^{2\pi}d\phi\int_{0}^{\pi} d\theta \sin \theta\int_{0}^{\pi}d\chi \sin^2\chi \int_{0}^{\infty}d\kappa \;\kappa^3\frac{{\rm e}^{-(\kappa^2 + m^2)/2\Lambda_{\kappa}^2}}{\kappa^2 + m^2}\\
    &= \frac{1}{8\pi^2}\int_{0}^{\infty}d\kappa \;\kappa^3\frac{{\rm e}^{-(\kappa^2 + m^2)/2\Lambda_{\kappa}^2}}{\kappa^2 + m^2}\\
    &= \frac{1}{8\pi^2}\left(\Lambda_p^2{\rm e}^{-m^2/2\Lambda_p^2} - m^2 Ei\left(1, m^2/2\Lambda_p^2\right)\right),
\end{align} where $Ei\left(1, m^2/2\Lambda_p^2\right)$ is the exponential integral. This expression is finite. Let us consider the ultra-relativistic limit for large $\Lambda_p$ :

\begin{equation}
    \label{okay}
    \tilde{\Delta}_F(0) \approx \frac{\Lambda_p^2}{8\pi^2}.
\end{equation} This expression for the Feynman propagator at $(x - x') = 0$ is finite in the limit $p^2 >> m^2$, which corresponds to the high energy limit. This difference is notable when compared to local QFT, because the propagator does not lead to UV divergences~\cite{sred2}. Moreover, the propagator is explicitly expressed in terms of the parameter $\Lambda_p$. This parameter is associated with the entire function distribution $\mathcal{E}(p^2) = \exp\left(-(p^2+m^2)/2\Lambda_p^2\right)$; it describes the uncertainty in momentum space. Associated with the uncertainty in position space $\Lambda_x$, it can be related to $\Lambda_p$ by the Heisenberg uncertainty principle~\cite{M5} :

\begin{equation}
    \Lambda_p\Lambda_x \geq 1.
\end{equation}

\subsection{Entanglement entropy in NLQFT}

Let $\Omega$ be the entire space, let also $A$ and $\bar{A}$ be two complementary half-spaces, therefore, $\Omega = A \cup \bar{A}$. In quantum field theory, the entanglement entropy corresponds to the correlations between the vacuum fluctuations. We define it as the von Neumann entropy~\cite{calabrese}:

\begin{equation}
    S = -Tr(\rho_A \ln(\rho_A)) = -Tr(\rho_{\bar{A}} \ln(\rho_{\bar{A}}),
\end{equation} where $\rho_A = Tr_{\bar{A}}(\rho_{\Omega})$ is the reduced density matrix and we have traced on the complementary region $\bar{A}$.

We compute the entanglement entropy $S$ by using the replica trick with regards to scalar fields in their vacuum state. We consider the Euclidean action on an $n$-sheeted Riemann surface with one cut along the negative real axis. The theory is defined on a cone with deficit angle $\delta = 2\pi (1 - n)$. We denote by $Z_n$ the partition function on the $n$-sheeted geometry. In order to compute the partition function, we use perturbation theory in powers of the coupling $\lambda$. Therefore, we can rewrite the entanglement entropy as a sum of the contributions of each order of the expansion. We have

\begin{equation}
    S = \sum_{k=0}^{+\infty}S_k,
\end{equation} where

\begin{equation}
    S_k = -\frac{\partial}{\partial n}(\ln Z_{n,k} - n \ln Z_{k,k})_{n \to 1}.
\end{equation}

Before starting the computation of the partition function of the Euclidean action, let us consider the two-point correlation function on the cone. It is the nonlocal Green's function that satisfies the following equation :

\begin{equation}
    (-\nabla^2_{\text{x}} + m^2)\tilde{G}_n(\text{x},\text{x}') = \mathcal{F}(\mathcal{E}(\text{x}))\delta^3(\text{x} - \text{x}'),
\end{equation} where $\nabla^2_{\text{x}}$ is the Laplacian differential operator.

The two-point correlation function on the cone breaks translation invariance. Therefore, it is a function of x and x$'$ separately instead of a function of $|\text{x} - \text{x}'|$ as in flat space. As in~\cite{H1, H2}, we write our coordinates the following way: x $=\{r, \theta, x\}$, where $\{r, \theta\}$ are the polar coordinates on the cone and $\{x\}$ are the coordinates on the transverse space. In this case, we work in $3+1$ dimensions, so the dimension of the transverse space is $2$. The nonlocal Green's function on the cone is given by~\cite{calabrese}:
\begin{equation}
    \tilde{G}_n(\text{x},\text{x}') = \int\frac{d^2p}{(2\pi)^2}\left[\sum_{k=0}^{+\infty}d_k\int_0^{\infty} \frac{dq}{2\pi n}\; q \frac{J_{k/n}(qr)J_{k/n}(qr')}{q^2 + p^2 + m^2}\cos(k(\theta - \theta')/n){\rm e}^{ip\cdot(x - x')}\right]{\rm e}^{\frac{-(p^2 + m^2)}{2\Lambda_p^2}},
\end{equation} where $J$ is the Bessel function of the first kind. The coefficients $d_k$ depend on $k$; $\forall k \geq 1, d_k = 2$ and $d_0$ = 1. Let us consider the coincident points, using the Euler-Maclaurin formula. The two-points correlation function simplifies to :

\begin{align}
    \begin{split}
        \tilde{G}_n(\text{x},\text{x}) = \frac{1}{2\pi n}\int\frac{d^2p}{(2\pi)^2}&\left[2\int_0^{\infty} dk \;  I_{k/n}(r\sqrt{p^2 + m^2})K_{k/n}(r\sqrt{p^2 + m^2}) \right. \\
        &\quad + \left.\frac{1}{6n}K_0^2(r\sqrt{p^2 + m^2})\right]{\rm e}^{-(p^2 + m^2)/2\Lambda_p^2},
    \end{split}
\end{align} where $I$ is the modified Bessel function of the first kind and $K$ is the modified Bessel function of the second kind.

The use of the Euler-Maclaurin formula allows to re-express the sum over the index $k$ in terms of an integral over $k$ and a sum over a dummy index $j$. For $j > 1$, the sum contribution does not lead to divergences in the local QFT. Therefore, we forget this sum.
For practical reasons, we rewrite

\begin{equation}
    \tilde{G}_n(\text{x},\text{x}) = \tilde{G}(\text{x},\text{x}) + \tilde{f}_n(r),
\end{equation} where $\tilde{G}(\text{x},\text{x})$ is the flat space Green's function at coincident points and $\tilde{f}_n(r)$ is given by

\begin{equation}
    \tilde{f}_n(r) = \frac{1}{2\pi n}\frac{1 - n^2}{6n}\int \frac{d^2p}{(2\pi)^2}K_0^2(r\sqrt{p^2 + m^2}){\rm e}^{-(p^2 + m^2)/2\Lambda_p^2}.
\end{equation} By translation invariance, $\tilde{G}(\text{x},\text{x}) = \tilde{G}(x,x) = \tilde{G}(0)$. Therefore, we rewrite

\begin{equation}
    \label{49}
    \tilde{G}_n(\text{x},\text{x}) = \tilde{G}(0) + \tilde{f}_n(r).
\end{equation}

In order to compute the entanglement entropy in the scalar field version of the NLQFT, we will limit ourselves to first order in the expansion of $\lambda$. Therefore, the partition function is given by

\begin{align}
    \ln Z_n &= \ln \int \mathcal{D}\phi \; {\rm e}^{-S_E[\phi]}\\
    &= \ln Z_{n,0} - \ln Z_{n,1}\\
    &= \ln Z_{n,0} - \frac{\lambda}{4!}\int_n d^4x \tilde{\phi}^4(x)\\
    &= \ln Z_{n,0} - \frac{3\lambda}{4!}\int_n d^4x \tilde{G}_n^2(x, x),
\end{align} where $Z_{n,0}$ is the partition function of the massive free scalar theory. The $3$-factor is a consequence of the application of Wick's theorem. By using the equation \eqref{49}, we have

\begin{equation}
    \ln Z_n = \ln Z_{n,0} - \frac{3\lambda}{4!}\int_n d^4x (\tilde{G}^2(0) + 2\tilde{G}(0)\tilde{f}_n(r) + \tilde{f}_n^2(r)).
\end{equation} To get the $S_1$ contribution for the entanglement entropy, such as defined by the replica trick,  we have to subtract $n$ times the partition function $Z_{1,1}$ :

\begin{align}
\label{replica}
    \ln Z_{n,1} - n\ln Z_{1,1} &= - \frac{3\lambda}{4!}\left[\int_n d^4x (\tilde{G}^2(0) + 2\tilde{G}(0)\tilde{f}_n(r) + \tilde{f}_n^2(r)) -n\int d^4x \tilde{G}^2(0)\right] \\
    &= - \frac{3\lambda}{4!}\left[\int_n d^4x (2\tilde{G}(0)\tilde{f}_n(r) + \tilde{f}_n^2(r))\right] \\
    &= - \frac{3\lambda}{4!}2\pi n \left[\int d^2x\int_0^{\infty} dr \;r (2\tilde{G}(0)\tilde{f}_n(r) + \tilde{f}_n^2(r))\right] \\
    \begin{split}
    &= - \frac{3\lambda}{4!}2\pi n
    A\left[\frac{2\tilde{G}(0)}{2\pi n}\frac{1 - n^2}{6n}\int\frac{d^2p}{(2\pi)^2}\int_0^{\infty}dr \;r K_0^2(r\sqrt{p^2 + m^2}){\rm e}^{\frac{-(p^2 + m^2)}{2\Lambda_p^2}} \right. \\
    &\quad \left. \quad + \int_0^{\infty} dr \;r  \tilde{f}_n^2(r)\right]
    \end{split}\\
    &= - \frac{3\lambda}{4!}A\left[\tilde{G}(0)\frac{1 - n^2}{6n}\int\frac{d^2p}{(2\pi)^2}\frac{{\rm e}^{-(p^2 + m^2)/2\Lambda_p^2}}{p^2 + m^2} + 2\pi n\int_0^{\infty} dr \;r  \tilde{f}_n^2(r)\right],
\end{align} where $A = \int d^2x$ is the area  resulting from the integration over the transverse space.

The last step to get the expression for $S_1$ is to differentiate with respect to $n$ and take the limit $n \to 1$. The second integral vanishes and we are left with the following derivation :

 \begin{equation}
     - \frac{\partial}{\partial n}\left(\frac{1 - n^2}{6n}\right)_{n \to 1} = \frac{1}{3}.
 \end{equation} Finally the $S_1$ term is given by

 \begin{equation}
     S_1 = - \frac{\lambda}{4!}A\tilde{G}(0)\int\frac{d^2p}{(2\pi)^2}\frac{{\rm e}^{-(p^2 + m^2)/2\Lambda_p^2}}{p^2 + m^2}.
 \end{equation} The solution for the free term is given by

\begin{equation}
    \ln Z_{n,0} = -\frac{1}{2}\ln \det(-\nabla^2_{\text{x}} + m^2).
\end{equation} We can link this quantity with the two-point correlation function by using its derivative with respect to $m^2$. We obtain :

\begin{equation}
    \frac{\partial}{\partial m^2} \ln Z_{n,0} = -\frac{1}{2} \int_n d^4x  \tilde{G}_n(\text{x},\text{x}).
\end{equation} To get $S_0$, we proceed by analogy with (\ref{replica}) :

\begin{align}
    \frac{\partial}{\partial m^2}(\ln Z_{n,0} - n\ln Z_{1,0}) &= - \frac{1}{2}\left[\int_n d^4x (\tilde{G}(\text{x},\text{x}) -n\int d^4x \tilde{G}(0)\right] \\
    &= - \frac{1}{2}\frac{1 - n^2}{6n}\int d^2x\int\frac{d^2p}{(2\pi)^2}\int_0^{\infty}dr \;r K_0^2(r\sqrt{p^2 + m^2}){\rm e}^{-(p^2 + m^2)/2\Lambda_p^2}  \\
    &= - A\frac{1 - n^2}{24n}\int \frac{d^2p}{(2\pi)^2}\frac{1}{p^2 + m^2}{\rm e}^{-(p^2 + m^2)/2\Lambda_p^2},
\end{align} using the fact that $\int_0^{\infty}dr \;r K_0^2(r\sqrt{p^2 + m^2}) = \frac{1}{2}\frac{1}{p^2 + m^2}$.

To get the $S_0$ term, we have to differentiate with respect to $n$ and to integrate with respect to $m^2$. The integration with respect to $m^2$ also gives a constant term independent of the mass $m^2$. In the local theory, this term is UV divergent. The UV divergence does not appear in the NLQFT due to the nonlocal regularization. In the following, we do not consider this constant term. Finally $S_0$ is given by

\begin{equation}
    S_0 = - \frac{A}{12}\int \frac{d^2p}{(2\pi)^2}\ln(p^2 + m^2){\rm e}^{-(p^2 + m^2)/2\Lambda_p^2}.
\end{equation}
The expression for the entanglement entropy for the scalar nonlocal field theory, in this configuration and at the first order of the coupling expansion, is given by
 \begin{align}
     S &= - \frac{A}{12}\int \frac{d^2p}{(2\pi)^2}\left[\ln(p^2 + m^2) + \frac{\lambda}{2} \frac{\tilde{G}(0)}{p^2 + m^2}\right]{\rm e}^{-(p^2 + m^2)/2\Lambda_p^2} + \mathcal{O}(\lambda^2) \\
     &= -\frac{A}{48\pi}\left(Ei(1, m^2/2\Lambda_p^2)\left(\frac{\lambda\tilde{G}(0)}{4} + \Lambda_p^2\right) + \Lambda_p^2 \ln(m^2){\rm e}^{-m^2/2\Lambda_p^2}\right) + \mathcal{O}(\lambda^2).
 \end{align} Note that because of the convergence of the nonlocal Feynman propagator at coincident points, the two-point correlation function at coincident points $\tilde{G}(0)$ is not divergent. Therefore, this expression for the entanglement entropy computed in $3+1$ dimensions on the cone is finite.

 This is a significant difference with the usual scalar field theory. Indeed, as discussed in~\cite{H2, cas1, cas2,agullo}, for such local scalar field theories, the entanglement entropy is a UV divergent quantity and this result is true for any manifold without boundary. Not only the term $S_0$ is logarithmically divergent but the term $S_1$ is also divergent, because the two-point correlation function at coincident points is quadratically divergent. In the local theory, it is therefore necessary to introduce a cutoff $\epsilon$ in order to regulate the divergences and extract a finite quantity from the entanglement entropy.

 In this section, we computed the entanglement entropy using the partition function which is based on the Lagrangian formalism. Therefore, we preserved all the symmetries of the theory. Because NLQFT is Lorentz invariant and causal, our result for the entanglement entropy is too. Our result has the advantage of being Lorentz invariant and finite. These two facts are consequences of nonlocal regularization using entire function distribution operators. The nonlocality taken as a hypothesis to build the theory makes it possible to obtain satisfactory results in the case that we have considered, that is to say a scalar field theory. In this nonlocal QFT, quantum correlations between entangled systems, i.e quantum entanglement, are direct consequences of the causal nonlocal assumption on which the theory is built.

 An interesting result is that the area law of entanglement entropy is preserved for this nonlocal QFT~\cite{sred1, H2, nishioka}.

\newpage
\section{Conclusions}

The field theory we consider is built from nonlocal field operators, the field at one point of spacetime depending on all other positions in spacetime. By using entire function distribution operators, our theory is made to be finite, which means there is no need for infinite renormalization, because the theory is free of UV-divergences. The nonlocality aims to reconcile the approach of relativistic field theory with the nonlocality observed experimentally on entangled quantum systems. The nonlocal QFT satisfies microcausality and no signals faster than light can be transmitted between two spacelike separated events.

The nonlocality results in the presence of an entire function distribution operator in the theory ${\cal E}=\exp(-(p^2-m^2)/2\Lambda_p^2)$, making it possible to obtain a finite partition function and entanglement entropy.

By considering the usual scalar field theory, we will always find that the entanglement entropy is UV divergent. Nonlocality in NLQFT makes it possible to obtain finite results, preserving Lorentz invariance and causality. We interpret quantum correlations between entangled systems as a direct consequences of nonlocality.

A similar study can be conducted for fermionic fields. The problem of infinite renormalizability is resolved by regulated propagators and vertices for the Feynman loop diagrams of QED, QCD and weak interactions~\cite{M5,M6}, and we expect to obtain finite entanglement entropies by considering configurations similar to the one presented in this paper.

It would be interesting to consider the framework of AQFT extended to nonlocal quantum field theory~\cite{AQFT}. AQFT would therefore be a nonlocal theory in the sense of operator algebras. The Reeh-Schlieder theorem will not be modified in the nonlocal QFT formulation.

\section{Data Availability Statement}

No datasets were generated or analyzed during the current study.

\section*{Acknowledgment}

We thank Laurent Freidel, Ivan Agullo, Viktor Toth and Martin Green for helpful discussions. This research was supported in part by Perimeter Institute for Theoretical Physics. Research at Perimeter Institute is supported by the Government of Canada through the Department of Innovation, Science and Economic Development Canada and by the Province of Ontario through the Ministry of Research, Innovation and Science.

\end{document}